\documentclass[aps,prb,reprint]{revtex4-2}
\usepackage{graphicx}
\usepackage{graphics}
\usepackage{amsfonts}
\usepackage{amsmath}
\usepackage{amssymb}
\usepackage{bm}

\begin{document}

\title{Numeric simulations build a bridge from a two-slit experiment \\
to the basics of X-ray diffraction and coherent optics}

\author{Ya.\ B. Bazaliy}
  \email{bazaliy@mailbox.sc.edu}
  \affiliation{University of South Carolina, Columbia, SC 29208, USA}

\date{\today}

\begin{abstract}
Numeric simulations based on the Huygens-Fresnel method allow one to develop intuition about the behavior of coherent light in diffraction and interference experiments. They give an opportunity to numerically observe and appreciate a number important phenomena, while avoiding the need to deal with the intricacies of their analytic descriptions. In an introductory teaching lab, they help to build a matrix of ideas, into which many optical demonstration experiments fall nicely and to the benefit of a student.
\end{abstract}

\maketitle

\section{Introduction}
Diffraction and interference are important and fascinating subjects encountered in many physics courses. At the same time, they are harder to learn because of the lack of practical familiarity: we do not have a chance to observe the behavior of coherent light in our everyday experiences. In that sense, there exists an noticeable difference between learning coherent optics and, say, introductory mechanics: in the latter case we can immediately compare the predictions of the theory with our existing intuitive understanding of mechanical motion.

Here we suggest that the lack of practical familiarity can be compensated by numeric simulations of coherent light based on the Huygens-Fresnel approach. The principles of the Huygens-Fresnel theory (Ref.~\onlinecite{hecht_book}, Sec. 10.1) are easy to explain, and usually are readily accepted by the students. However, to implement them one needs to use sophisticated methods of calculus, involving difficult integrations that sometimes result in special functions, and to consider infinite sums converging to delta functions. Those technicalities obscure the simple foundations of the method and dramatically discourage many students.

Resorting to numerics allows one to lead the students to discover for themselves the notions that otherwise have to be introduced without sufficient motivation. Numerical experiments allow one to change parameters that are not easily tunable in real-life optical measurements and thus smoothly connect the results of a number of impressive qualitative demonstrations that otherwise may seem to be quite disjoint. Overall, computation can and should play a meaningful role in the teaching of diffraction and interference phenomena.

Numeric calculations may be also viewed as a substitute for the method of {\em phasor diagrams} (see Ref.~\onlinecite{hecht_book}, Sec.~9.6, 10.2, 10.3). The latter are designed to visualize the summation, or integration, of the secondary source contributions in the Huygens-Fresnel approach. When mastered, the phasor method provides a semi-analytic way to understand the diffraction and interference phenomena. Numeric summation, on the other hand, may be used to achieve the same goal when students are not expected to spend that much time on the subject.

The paper is organized to be as self-contained as possible. It partially repeats the textbook material, adapting it so as to support the numeric examples. This is done with the goal of making it handier for potential instructors and students alike. Implementation of the simulation code in the form of a {\em Mathematica} notebook is provided in the Supplement.

\section{Huygens-Fresnel summations}\label{sec:H-F}
\subsection{Basic case of two infinitely thin slits}

\begin{figure}[t]
\center
\includegraphics[width = 0.35\textwidth]{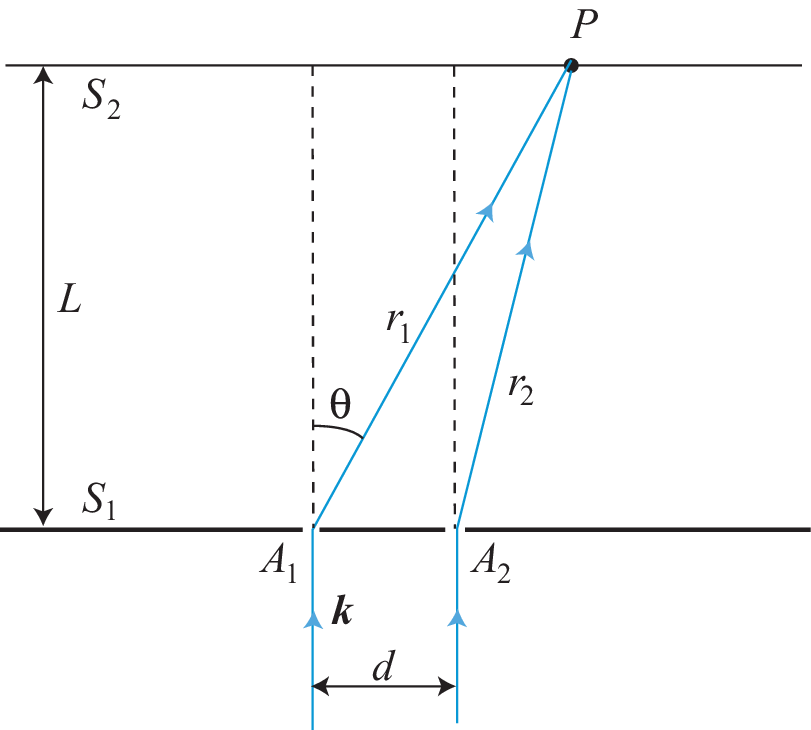}
\caption{Two-slit interference with finite distance to the screen}
 \label{fig:two_slits_1}
\end{figure}

We start with a two-slit interference setup shown in Fig.~\ref{fig:two_slits_1}. The incoming plane wave of light goes through two infinitely thin slits $A_1$ and $A_2$ located a distance $d$ apart on the screen $S_1$. Interference pattern is observed on the screen $S_2$. Screens are parallel to each other with a distance $L$ between them. The wave vector $\bf k$ of the incoming light is perpendicular to the screens. Interference is observed at a point $P$ on $S_2$. According to the Huygens-Fresnel principle, each point-like slit is a source of a spherical wave. From the very beginning we will employ the complex notation. Then the wave's amplitudes at $P$ are given by the expressions
\begin{equation}\label{eq:spherical_wave}
a_i(P) = a_0 \frac{e^{i k r_i}}{r_i}, \qquad i = 1,2
\end{equation}
where $r_i$ are the distances $|A_iP|$, $k = |{\bf k}|$, and $a_0$ is the amplitude produced by the incoming light at the plane of screen $S_1$ (same for both slits). Intensity $I_2$ of the light produced by two slits at $P$ is then found as the absolute value of total amplitude squared
\begin{eqnarray}
\nonumber
I_2(P) &=& |a_1 + a_2|^2 = |a_0|^2 \left| \frac{e^{i k r_1}}{r_1} + \frac{e^{i k r_2}}{r_2} \right|^2
\\
\label{eq:IP2}
&=& |a_0|^2 \left|
  \frac{1}{r_1} + \frac{e^{i k \Delta l}}{r_2}
  \right|^2 \ .
\end{eqnarray}
with $\Delta l = r_2 - r_1$ being the difference in optical paths. Expression (\ref{eq:IP2}) can be easily evaluated numerically but here we will further specialize to the {\em Fraunhofer diffraction} regime (Ref.~\onlinecite{hecht_book}, Sec. 10.1.2) that occurs when the distance between the screens is large enough to satisfy the ineqality
\begin{equation}\label{eq:Fraunhofer_diffraction_condition}
k \frac{d^2}{L} \ll 1 \ .
\end{equation}
(Nice numeric comparison of different diffraction regimes can be found in Ref.~\onlinecite{Davidovic:PT2019:Fraunhofer-Fresnel}). It is well known \cite{hecht_book} that in that case the intensity (\ref{eq:IP2}) can be approximated by
\begin{equation}\label{eq:Fraunhofer_intensity}
I_2(\theta) \approx \frac{|a_0|^2}{r^2} \left|
1 + e^{-i k d \sin\theta}
\right|^2 \ ,
\end{equation}
where the angle $\theta$ is defined in Fig.~\ref{fig:two_slits_1}.

We are not going to re-derive the Fraunhofer approximation (\ref{eq:Fraunhofer_intensity}) but wish to make a remark. In a naive way, Eq.~(\ref{eq:Fraunhofer_intensity}) may be obtained from Eq.~(\ref{eq:IP2}) by setting $\Delta l = 0$ in the denominators and writing $r_1 = r_2 = r$, while at the same time using the expression $\Delta l = - d \sin\theta$ in the exponent. Such procedure may seem inconsistent at first but in fact results from a rigorous process of Taylor expansion of (\ref{eq:IP2}) in the limit specified by the inequality (\ref{eq:Fraunhofer_diffraction_condition}). The necessity of different treatments of $\Delta l$ it two positions comes from the fact that the pro\-duct $k \Delta l = 2\pi \Delta l/\lambda$ can be large compared to unity even when $\Delta l$ itself is small compared to $r$. This happens because the wavelength of light $\lambda$ is the smallest length scale in the problem with $\lambda \ll \Delta l$.

\begin{figure}[t]
\center
\includegraphics[width = 0.35\textwidth]{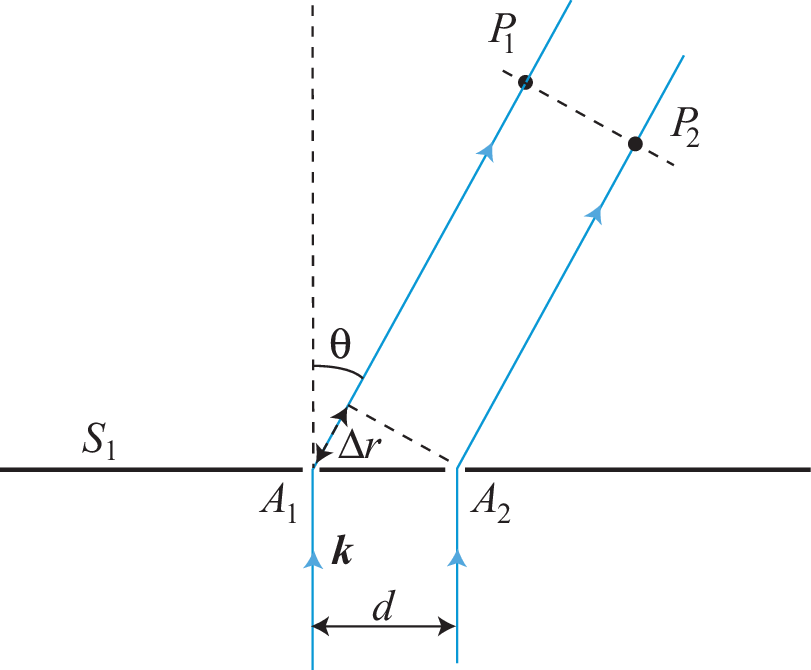}
\caption{Two-slit interference for a screen ``at infinity''}
 \label{fig:two_slits_2}
\end{figure}

\begin{figure}[b]
\center
\includegraphics[width = 0.35\textwidth]{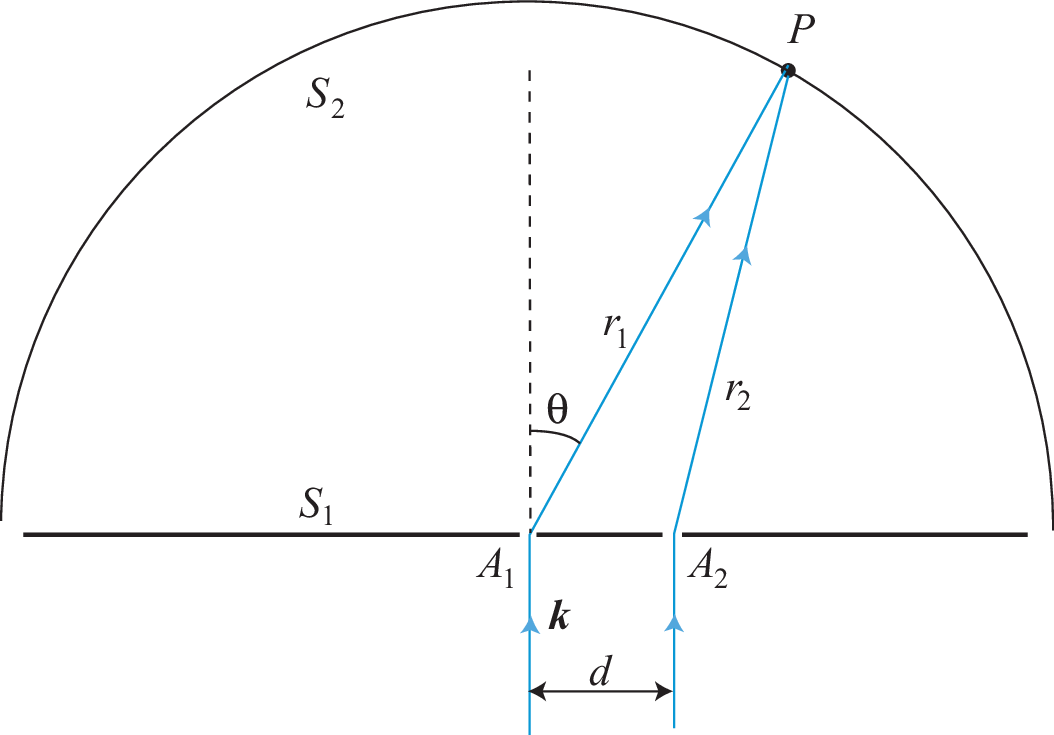}
\caption{Two-slit interference on a far away circular screen}
 \label{fig:two_slits_3}
\end{figure}

Formula (\ref{eq:Fraunhofer_intensity}) is often illustrated by Fig.~\ref{fig:two_slits_2}, showing the limit of ``diffraction at infinity''. The intervals $A_1P$ and $A_2P$ from Fig.~\ref{fig:two_slits_1} are replaced here by the parallel rays $A_1P_1$ and $A_2P_2$, with $P_1 P_2$ being perpendicular to the rays. Optical paths difference in the exponent is given by $\Delta l = |A_2P_2| - |A_1P_1|$, and it is easy to see from Fig.~\ref{fig:two_slits_2} that one gets $\Delta l = - d \sin\theta$. Figure~\ref{fig:two_slits_2} essentially represents the limit reached by Fig.~\ref{fig:two_slits_1} when the angle $\angle A_1PA_2$ goes to zero. Before the limit is taken, points $P_1$ and $P_2$ are defined as located at the same distances from the far away point $P$; the limiting procedure makes the rays going to $A_1P$ and $A_2P$ parallel, and the interval $P_1 P_2$ perpendicular to them.

When the angle $\theta$ is small, $r = L/\cos\theta \approx L$ and we can re-write (\ref{eq:Fraunhofer_intensity}) as
\begin{equation}\label{eq:Fraunhofer_intensity-2}
I_2(\theta) = I_0 \left|
1 + e^{-i k d \sin\theta}
\right|^2 \ , \qquad
I_0 = \frac{|a_0|^2}{L^2} \ .
\end{equation}

One can further imagine an experiment with a circular screen $S_2$ of radius $L$, as shown in Fig.~\ref{fig:two_slits_3}. In such setting there will be no restrictions on $\theta$ in formula (\ref{eq:Fraunhofer_intensity-2}).

\subsection{Numeric result for the two-slit interference}\label{sec:2slit_w=0}
\begin{figure}[h]
\center
\includegraphics[width = 0.3\textwidth]{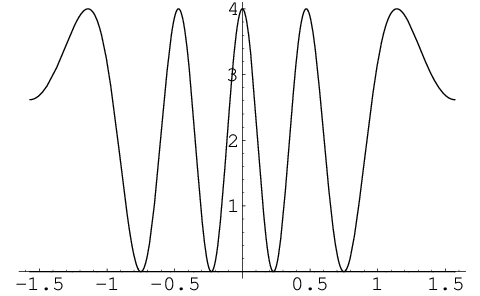}
\caption{Two-slit interference pattern $I_2(\theta)/I_0$ for $d = 2.2 \lambda$}
 \label{fig:two_slits_numeric_1}
\end{figure}

In the spirit of our approach, we do not try to work analytically even with the simplest formula (\ref{eq:Fraunhofer_intensity-2}). Instead, the students are asked to code this formula and plot the results. Figure~\ref{fig:two_slits_numeric_1} shows a representative plot of $I_2(\theta)$. The students are instructed to experiment with the value of $d$, expressed in terms on $\lambda$. They observe how the increase of the $d/\lambda$ ratio produces more peaks in the interference picture.

\subsection{Diffraction grating made of equally spaced, infinitely thin slits}\label{sec:Nslit_w=0}

Generalizations of the two-slit formula (\ref{eq:Fraunhofer_intensity-2}) for a larger number of slits, i.e., to a diffraction grating, are intuitively believable. For example, for three slits, equally spaced by the distances $d$
$$
I_3(\theta) = I_0 \left| 1 + e^{-i k d \sin\theta} + e^{-2 i k d \sin\theta} \right|^2 \ .
$$
In this formula the last term has a form of $\exp(i k \Delta l_2)$, with $\Delta l_2 = - 2 d \sin\theta$ being the optical path difference between the third and the first slits. Expression for $I_3$ can be derived by either following the procedure that led from (\ref{eq:IP2}) to (\ref{eq:Fraunhofer_intensity-2}), or by generalizing Fig.~\ref{fig:two_slits_2} to the three-slit case.

\begin{figure}[t]
\center
\includegraphics[width = 0.4\textwidth]{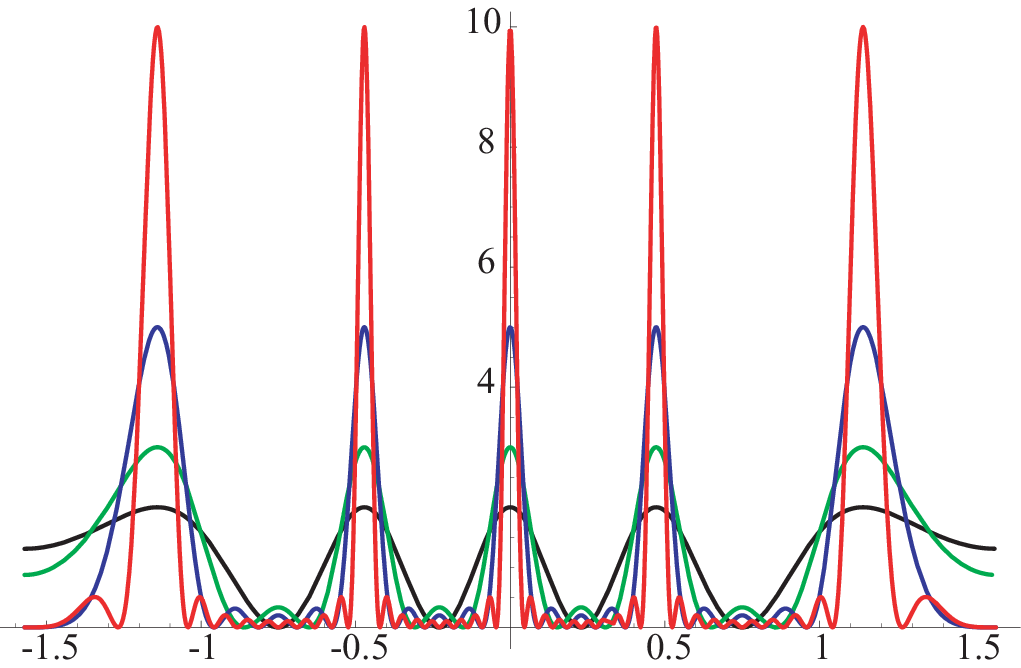}
\caption{Grating with infinitely thin slits. Plots of $I_N(\theta)/(N I_0)$ for $d = 2.2 \lambda$ and a series of $N$ values. Black: $N = 2$. Green: $N = 3$. Blue: $N = 5$. Red: $N = 10$}
 \label{fig:many_slits_numeric}
\end{figure}

For an arbitrary number $N$ of equally spaced, infinitely thin slits, reads (Ref.~\onlinecite{hecht_book}, Sec. 10.1.3)
\begin{equation}\label{eq:Fraunhofer_intensity_N_slits}
I_N(\theta) = I_0
\left|
   \sum_{p = 0}^{N-1} e^{-i (k d \sin\theta) p}
\right|^2 \ .
\end{equation}
Plotting this formula for several $N$ values in Fig.~\ref{fig:many_slits_numeric} clearly demonstrates the main qualitative result: for a given $d$, the increase of $N$ makes each intensity peak of the two-slit experiment progressively higher and narrower, while the angular positions of the peaks do not change. Multiple smaller peaks are produced but those are much lower than the ``major'' ones.

The intuitive picture acquired by the students is of the ``squeezing'' of the light into the progressively narrower beams. As a result of such squeezing the intensity of each beam goes up.

Note that the intensity given by formula (\ref{eq:Fraunhofer_intensity_N_slits}) grows with $N$ for two reasons. The first one is trivial: more slits mean that a larger amount of incident light is transmitted by the $S_1$ screen. The second one is that the squeezing effect redistributes the light over the angles. In order to exclude the trivial effect, Fig.~\ref{fig:many_slits_numeric} shows the plots of the normalized quantity $I_N(\theta)/(N I_0)$. One can clearly see from the graph that at the major maxima the normalized intensity reaches the value of $I_N(\theta_{max})/(N I_0) = N$. It is a little more difficult task to check numerically that the widths of the main maxima are proportional to $1/N$.

After getting familiar with the behavior of major peaks, the students are encouraged to check numerically that $I_N$ can be alternatively obtained from the analytic formula (Ref.~\onlinecite{hecht_book}, Sec. 10.1.3)
\begin{equation}\label{eq:formula_for_sum_of_exponents}
I_N(\theta) = I_0 \left[
 \frac{\sin (N [k d \sin\theta]/2)}{\sin ([k d \sin\theta]/2)}
\right]^2 \ .
\end{equation}

\subsection{Single slit of finite width}\label{sec:1slit_w}
Real-life slits have finite width $w$, and the next numeric exercise addresses this issue. In the Huygens-Fresnel approach a finite size slit should be represented by a collection of an infinite number of infinitely weak secondary sources. In other words, in an analytical approach it is given by an integral.

Numerically, the best we can do is to approximate that integral by sums. We divide the slit into a large enough number $M$ of secondary sources and calculate the sum
\begin{equation}\label{eq:Fraunhofer_intensity_finite-width_slit}
J_M(\theta) = J_0^{(M)}
\left|
   \sum_{p = 0}^{M-1} e^{-i (k [w/M] \sin\theta) p}
\right|^2 \ ,
\end{equation}
where $J_0^{(M)}$ is the intensity of each secondary source, a quantity to be discussed below.

Expressions (\ref{eq:Fraunhofer_intensity_finite-width_slit}) and (\ref{eq:Fraunhofer_intensity_N_slits}) look very similar, however, there is a crucial difference. In Eq.~(\ref{eq:Fraunhofer_intensity_N_slits}), as $N$ is increased the intensity of a single secondary source and the distance $d$ between the sources remains the same, while the overall intensity and the overall width of the grating grow indefinitely. In Eq.~(\ref{eq:Fraunhofer_intensity_finite-width_slit}) the intensity of each source $I_0^{(M)}$ and the distance $d_M = w/M$ between the sources decrease with $M$, while the overall width of the slit remains constant and the overall intensity approaches a certain limit.

When choosing the secondary source intensity $J_0^{(M)}$ in Eq.~(\ref{eq:Fraunhofer_intensity_finite-width_slit}), we face a dilemma. Clearly, the total amount of light falling on the slit has to be divided by the number $M$ of the secondary sources. But should one divide the amplitude or the intensity of the incoming light? The problem is that the two choices are incompatible. If one chooses to divide the amplitude and write $a_0^{(M)} = a/M$, then the intensity is given by $J_0^{(M)} = |a_0^{(M)}/L|^2 = |a|^2/L^2 M^2 \sim 1/M^2$. But if one chooses to divide the intensity, the relationship $J_0^{(M)} \sim 1/M$ will hold.

The correct choice turns out to be the division of amplitude (Ref.~\onlinecite{hecht_book}, Sec. 10.1.3), and formula (\ref{eq:Fraunhofer_intensity_finite-width_slit}) acquires a form
$$
J_M(\theta) = \frac{|a|^2}{L^2 M^2}
\left|
   \sum_{p = 0}^{M-1} e^{-i (k [w/M] \sin\theta) p}
\right|^2 ,
$$
where parameter $|a|^2$ is ultimately determined by the characteristics of the incoming light. It is beyond the scope of our paper to derive $|a|^2$. Fortunately, if we are only interested in the shape of the diffraction pattern and are not concerned with its overall intensity, we can avoid that difficult task. Let's relate $|a|^2$ to the diffraction intensity at $\theta = 0$. This gives $J_M(0) = |a|^2/L^2$, and so
\begin{equation}\label{eq:Fraunhofer_intensity_finite-width_slit-2}
J_M(\theta) = \frac{J_M(0)}{M^2}
\left|
   \sum_{p = 0}^{M-1} e^{-i (k [w/M] \sin\theta) p}
\right|^2 ,
\end{equation}

\begin{figure}[t]
\includegraphics[width = 0.45\textwidth]{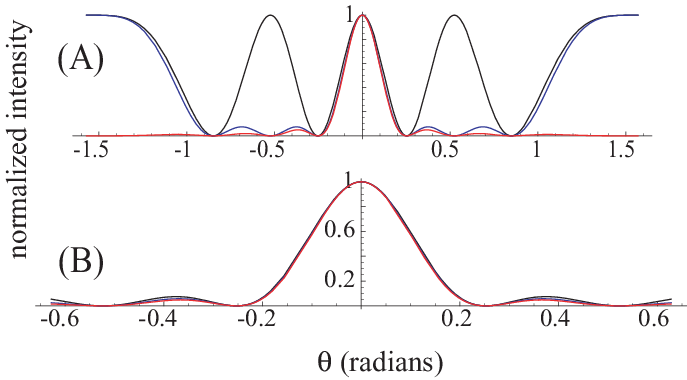}
\caption{Normalized diffraction intensity $J_M(\theta)/J_M(0)$ for a single finite-width slit of width $w = 4 \lambda$. Figure shows how approximations (\ref{eq:Fraunhofer_intensity_finite-width_slit-2}) converge as $d_M = w/M$ decreases. (A) View of the full angular range with black line for $d_M/\lambda = 2$, blue line for $d_M/\lambda = 1$, red line for $d_M/\lambda = 0.33$. (B) Comparison of numeric results with the analytic formula in the cental peak area. Black line for $d_M/\lambda = 1$, blue line for $d_M/\lambda = 0.66$, red line given by formula (\ref{eq:Fraunhofer_intensity_finite-width_analytic}). }
 \label{fig:finite-width_slit_numeric}
\end{figure}

Students are asked to plot a series of $J_M(\theta)$ dependencies (\ref{eq:Fraunhofer_intensity_finite-width_slit-2}) with increasing values of $M$. Representative plots are shown in Fig.~\ref{fig:finite-width_slit_numeric}(A). Numeric graphs rapidly converge to a limiting function $J_M(\theta) \to J(\theta)$ as soon as $d_M \lesssim \lambda$ is achieved. One can also check that the limit is described by the analytic formula (Ref.~\onlinecite{hecht_book}, Sec. 10.2.1)
\begin{equation}\label{eq:Fraunhofer_intensity_finite-width_analytic}
J(\theta) = J(0) \left[
  \frac{\sin \big( [k w \sin\theta]/2 \big)}{[k w \sin\theta]/2}
\right]^2 ,
\end{equation}
as illustrated by Fig.~\ref{fig:finite-width_slit_numeric}(B).

The difference between Fig.~\ref{fig:many_slits_numeric} and Fig.~\ref{fig:finite-width_slit_numeric} is worth a remark, as it prominently illustrates the general idea:  a limit of a mathematical expression with several parameters may depend on the path in the parameter space. In the case of many slits, we take a limit $N \to \infty$ but keep $d$ and $I_0$ constant. In the case of a single finite-width slit $d_M$ and $J_0^{(M)}$ are changed in concert with $M \to \infty$.

\subsection{Diffraction grating made of finite-width slits}\label{sec:form-factor}
A grating made of infinitely thin slits, and a finite-width single slit both exhibit diffraction maxima and minima. How do these features combine in the interference pattern produced by a real-life grating consisting of $N$ slits of finite width $w$ spaced by a period $d > w$?

\begin{figure}[b]
\center
\includegraphics[width = 0.4\textwidth]{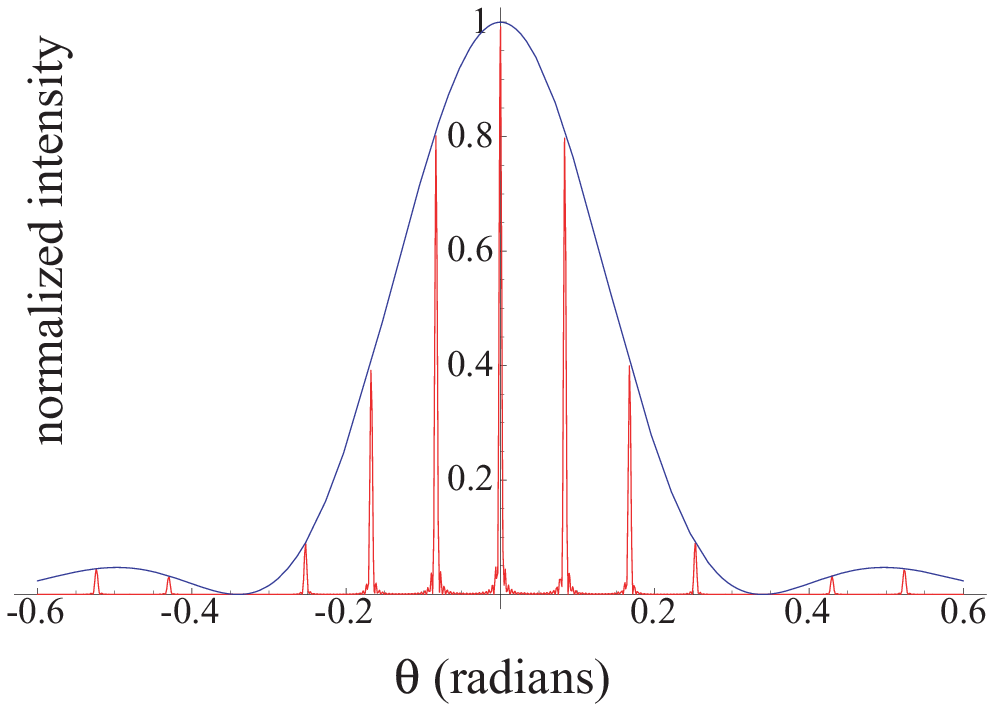}
\caption{Grating with finite-width slits. Red: normalized diffraction intensity $I(\theta)/J_M(0) N^2$ for a grating with $N = 20$ slits of width $w = 3 \lambda$ and a period $d = 12 \lambda$. Blue: normalized diffraction intensity $J_M(\theta)/J_M(0)$ on a single slit of width $w = 3 \lambda$.}
 \label{fig:many_finite_width_slits_numeric}
\end{figure}

To investigate this issue in our numeric approach, we should divide every slit into $M \gtrsim w/\lambda$ secondary sources and then sum the $N \times M$ contributions. Each secondary source is labeled by a pair $(p,q)$, where $p$ is the number of the slit, $p = 1,2, \ldots , N$, and $q$ is the number of the source within a given slit, $q = 1,2, \ldots , M$. The optical path difference between the source $(p,q)$ and the source $(1,1)$ is then given by
\begin{equation}\label{eq:Delta_r}
\Delta l(p,q) = -\left[
d \sin\theta (p-1) + \frac{w}{M} \sin\theta(q-1)
\right]  \ .
\end{equation}
The sum of the secondary source waves propagating in the direction $\theta$ assumes the form
\begin{equation}\label{eq:many_finite-width_slits-1}
a(\theta) =  \sum_{p = 1}^{N} \sum_{q = 1}^{M}
\frac{a_0^{(M)}}{L} e^{i k \Delta l(p,q) } \ .
\end{equation}
Initially, the students are asked to directly write a code implementing  formula (\ref{eq:many_finite-width_slits-1}) and plotting the intensity $I(\theta) = |a(\theta)|^2$ for a grating of finite-size slits. This provides a graph (Fig.~\ref{fig:many_finite_width_slits_numeric}) that, similarly to the case of infinitely thin slits (Fig.~\ref{fig:many_slits_numeric}), features a set of sharp diffraction peaks. The crucial difference is that now the heights of the peaks are variable.

Next, the pattern in Fig.~\ref{fig:many_finite_width_slits_numeric} is used to motivate the students to make a useful analytic step. They are challenged to prove the factorization of the sum (\ref{eq:many_finite-width_slits-1}) into a product
\begin{eqnarray}
\nonumber
a &=& \frac{a_0^{(M)}}{L} \left(
  \sum_{p = 1}^{N} e^{-i [k d \sin\theta] (p-1)  }
  \right)
\\
\label{eq:amplitude_factorization}
 & \times & \left(
  \sum_{q = 1}^{M} e^{-i [  k (w/M) \sin\theta] (q-1)}
  \right)
\end{eqnarray}
If that is the first time a student encounters such factorization, it is well worth going through a detailed derivation and showing that it is the additive nature of $\Delta l(p,q)$ that leads to the factorization of $\exp(i k \Delta l(p,q))$, and ultimately enables the factorization in Eq.~(\ref{eq:amplitude_factorization}). An explicit counter-example of a double sum that cannot be factorized is a good idea. Demystifying the process now will help on multiple future occasions when similar factorizations are encountered in other physics courses.

With factorization (\ref{eq:amplitude_factorization}) the intensity can be written as
\begin{equation}\label{eq:intensity_factorization-1}
I =  \left|
  \sum_{p = 1}^{N} e^{-i [k d \sin\theta] (p-1)  }
  \right|^2 \times
  \left|
  \frac{a_0^{(M)}}{L}
  \sum_{q = 1}^{M} e^{-i [k (w/M) \sin\theta] (q-1) }
  \right|^2
\end{equation}
The second term is recognized as nothing else but the single slit intensity (\ref{eq:Fraunhofer_intensity_finite-width_slit}). Thus
\begin{equation} \label{eq:intensity_factorization-2}
I = \left|
  \sum_{p = 1}^{N} e^{-i [k d \sin\theta] (p-1) }
  \right|^2 J_M(\theta)
  = \left( \frac{I_N(\theta)}{I_0} \right) J_{M}(\theta)
\end{equation}
with $I_N$ given by formula (\ref{eq:Fraunhofer_intensity_N_slits}) and $J_M$ given by formula (\ref{eq:Fraunhofer_intensity_finite-width_slit-2}). Function $I_N(\theta)/I_0$ is called the {\em grating factor}. It reflects the periodicity and size of the grating. Function $ J_{M}(\theta)$ is called the {\em form-factor}. It is determined by one element of the grating, here a slit of width $w$.

Factorization (\ref{eq:intensity_factorization-2}) nicely explains the features of Fig.~\ref{fig:many_finite_width_slits_numeric}. The grating factor is responsible for the diffraction peaks with heights proportional to $N^2$. The form-factor modulates those peaks' heights. Figure~\ref{fig:many_finite_width_slits_numeric} shows the normalized intensity $I(\theta)/J_M(0) N^2$, for which $J_M(\theta)/J_M(0)$ serves as an envelope function.

The wider message of this section for the students is that the features of a diffraction pattern are in a way ``inverted'' as compared to the features of the grating itself. In real space, the slit size is a local property, and the number of slits is a global property of the grating. In the diffraction pattern the number of slits determines the local property, i.e., the width of individual peaks, while the global pattern formed by the relative peak intensities is determined by the characteristics of a single slit.

In order to get information about the local property of an individual slit, one has to study the global property of the diffraction pattern. Periodic arrangement of slits into a grating does not change this global property. The role of the  grating is to enhance the intensity of light in major peaks, making their heights easily measurable. Note that the heights of the peaks grow proportional to $N^2$, so the enhancement is much larger than one would get from just adding $N$ equal contributions. The price for such help is that the envelope is only measured on a set of discrete angles corresponding to peak positions, instead of being known as a continuous function of angle.

\section{Describing diffraction in terms of the reciprocal lattice}\label{sec:reciprocal_lattice}

In class demonstrations with diffraction gratings the bright diffraction peaks are the most prominent qualitative features noticed by the students. This section introduces the idea of a reciprocal lattice as the most useful tool for understanding the peak positions. Observing the modulation of peak intensities may be possible but requires more focused attention.

\subsection{Tilted diffraction grating}

\begin{figure}[h]
\center
\includegraphics[width = 0.3\textwidth]{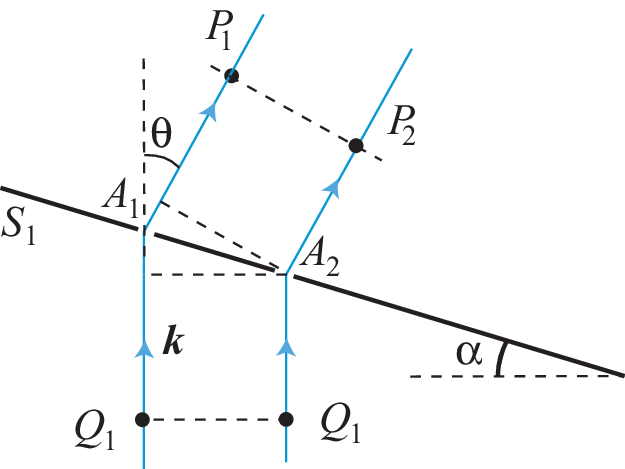}
\caption{Interference at infinity from two point sources on a tilted diffraction grating. The tilt angle $\alpha$ is considered positive if the screen $S_1$ is rotated clockwise.}
 \label{fig:two_slits_tilted}
\end{figure}

To motivate the discussion, we start with a demonstration experiment where peak positions can be con\-ti\-nuously tuned by tilting the grating (Fig.~\ref{fig:two_slits_tilted}). In the literature, diffraction on tilted gratings is shown to exhibit interesting non-monotonic dependence of peak angles on the tilt angle \cite{lock:PT1985:rotated_grating}.

We consider the same grating as in the previous section, except that now it is tilted with respect to the incoming light wave by an angle $\alpha$, as shown in Fig.~\ref{fig:two_slits_tilted}. Such situation can be easily achieved experimentally with an adjustable tilt angle. Note that in the case of tilted grating the diffraction angle $\theta$ changes in the interval $-\pi/2+\alpha \leq \theta \leq \pi/2 + \alpha$.

With a tilted grating, sources $A_1$ and $A_2$ are no longer in-phase. Consequently, the expression for the optical path difference changes to
$$
\Delta l = |Q_1A_1| + |A_1P_1| - |Q_2A_2| - |A_2P_2| \ .
$$
Geometric arguments based on Fig.~\ref{fig:two_slits_tilted} produce the formula
\begin{equation}\label{eq:tilted_path_difference}
\Delta l(\theta) = - d[\sin(\theta - \alpha) + \sin\alpha] = - d \, f(\theta,\alpha) \ .
\end{equation}
This result means that all expression derived for the perpendicular incoming beam can be re-used by simply substituting $f(\theta,\alpha)$ for $\sin\theta$ in the exponents. That substitution changes both the grating factor and the form factor.

\subsection{Positions of major peaks}
Our goal is to follow the positions of the major peaks as the tilting angle is gradually changed. Every textbook discussing diffraction on gratings shows that the positions of major maxima in Fig.~\ref{fig:many_finite_width_slits_numeric} can be obtained by requiring that all terms in the grating factor sum (\ref{eq:Fraunhofer_intensity_N_slits}) are equal to unity
$$
e^{-i (k d \sin\theta) p} = 1 \qquad {\rm for \ any} \ p \ ,
$$
or equivalently
$$
k d \sin\theta_m = 2 \pi m
$$
with an integer $m$. Figures.~\ref{fig:many_slits_numeric} and \ref{fig:many_finite_width_slits_numeric} can be used to numerically extract $\theta_m$'s and check this condition. For a tilted grating the requirement is obviously generalized to
\begin{equation}\label{eq:main_maxima_locations}
k d \, f(\theta_m,\alpha) = 2 \pi m
\end{equation}

\subsection{Limit of an infinitely large grating}
While formula (\ref{eq:main_maxima_locations}) for the locations of the major peaks can be justified by various plausible arguments, it is ultimately based on a mathematical statement about the grating factor expression
$$
\left|
  \sum_{p = 1}^{N} e^{-i x (p-1) }
  \right|^2 \ ,
$$
encountered in Eqs.~(\ref{eq:Fraunhofer_intensity_N_slits}) and (\ref{eq:intensity_factorization-2}) with $x = k d \, f(\theta)$. This expression has a few equivalent forms. Because of the absolute value involved, it can be represented as
$$
\left|
  \sum_{p = 1}^{N} e^{-i x (p-1) }
  \right|^2
= \left|
  \sum_{p = 1}^{N} e^{i x p }
  \right|^2 \ .
$$
Also, Eq.~(\ref{eq:formula_for_sum_of_exponents}) gives an explicit formula
$$
\left|
  \sum_{p = 1}^{N} e^{-i x (p-1) }
  \right|^2
= \left[
 \frac{\sin (N x/2)}{\sin (x/2)}
\right]^2 \ .
$$
Sharp diffraction peaks produced by formulas (\ref{eq:Fraunhofer_intensity_N_slits}) and (\ref{eq:intensity_factorization-2}) for the large values of $N$ result from the limiting property
\begin{equation}\label{eq:delta_function_limit}
 \frac{1}{N} \left|
  \sum_{p = 1}^{N} e^{-i x (p-1) }
  \right|^2 \xrightarrow[]{N \to \infty}
  2 \pi \sum_{m = -\infty}^{\infty} \delta(x - 2\pi m)
\end{equation}
where the sum of delta functions on the right hand side is taken over all integers $m$. The delta functions mean that intensities are non-zero only for $x = 2\pi m$, which is equivalent to the statement of Eq.~(\ref{eq:main_maxima_locations}).

A rigorous discussion of Eq.~(\ref{eq:delta_function_limit}) is beyond the level of calculus expected from the students attending an introductory physics course of the type discussed here. Nevertheless, numeric evaluations of the sums of exponents and their plots in Figs.~\ref{fig:many_slits_numeric} and \ref{fig:many_finite_width_slits_numeric} have already provided them with an empirical evidence of the reality of sharp peaks and build enough intuition to accept formula (\ref{eq:delta_function_limit}) without proof.

\subsection{Vector formula for the phase difference}\label{sec:vector_formula_for_phase}
To explain the positions of the peaks in demonstration experiments with gratings, one wants to find the angles $\theta_m$ of the major diffraction maxima, i.e., to invert equation (\ref{eq:main_maxima_locations}). Such inversion is not hard to do nume\-ri\-cally but here our goal will be to bring up a geometric construction that makes the picture much more transparent. Namely, we are going to introduce the {\em reciprocal lattice} associated with a given grating.

We start by re-deriving the function $f(\theta)$ using vector algebra instead of planar geometry. Consider Fig.~\ref{fig:two_slits_tilted} and denote the radius-vectors sources $A_1$ and $A_2$ as ${\bf R}_1$ and ${\bf R}_2$. Denote the radius-vectors of points $Q_{1,2}$ as ${\bf r}_{1,2}$, and denote the radius vectors of points $P_{1,2}$ as ${\bm\rho}_{1,2}$. Let the incoming light have wave vector ${\bf k}_{in}$ and the diffracted light have wave vector ${\bf k}_{out}$. The absolute values of the two are equal and denoted as $k$. Diffraction angle $\theta$ is measured between ${\bf k}_{in}$ and ${\bf k}_{out}$. Two parallel incoming rays reaching the sources are going along ${\bf k}_{in}$, and two parallel outgoing diffracted rays, originating from the sources, are going along ${\bf k}_{out}$. By construction, vector ${\bf r}_2 - {\bf r}_1$ is perpendicular to ${\bf k}_{in}$, and vector ${\bm \rho}_2 - {\bm \rho}_1$ is perpendicular to ${\bf k}_{out}$. In terms of the introduced vectors one can write
\begin{eqnarray*}
k \Delta l &=& \big( {\bf k}_{in} \cdot ({\bf R}_2 - {\bf r}_2) + {\bf k}_{out} \cdot ({\bm \rho}_2 - {\bf R}_2) \big)
\\
&-&   \big( {\bf k}_{in} \cdot ({\bf R}_1 - {\bf r}_1) + {\bf k}_{out} \cdot ({\bm \rho}_1 - {\bf R}_1) \big) \ .
\end{eqnarray*}
Collecting the terms and using the properties $({\bm \rho}_2 - {\bm \rho}_1) \cdot {\bf k}_{out} = 0$ and $({\bf r}_2 - {\bf r}_1) \cdot {\bf k}_{in} = 0$, one gets
$$
k \Delta l = - ( {\bf k}_{out} -  {\bf k}_{in} ) \cdot
({\bf R}_2 - {\bf R}_1) \ .
$$
Denoting
\begin{eqnarray}
\label{eq:scattering_vector}
{\bf q} &=& {\bf k}_{out} -  {\bf k}_{in} \ ,
\\
\label{eq:displacement_vector}
\delta{\bf R} &=& {\bf R}_2 - {\bf R}_1 \ ,
\end{eqnarray}
we reach the final formula
\begin{equation}\label{eq:phase_in_vector_notation}
k \Delta l = - {\bf q} \cdot \delta{\bf R} \ .
\end{equation}

\begin{figure}[b]
\center
\includegraphics[width = 0.17\textwidth]{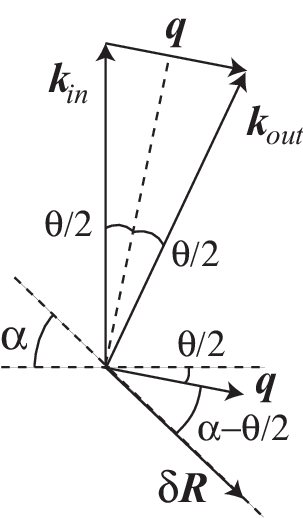}
\caption{Proving equivalence between Eqs.~(\ref{eq:phase_in_vector_notation}) and (\ref{eq:tilted_path_difference})}
 \label{fig:two_slits_qR}
\end{figure}

Figure~\ref{fig:two_slits_qR} shows that in the case of a tilted grating formulas (\ref{eq:phase_in_vector_notation}) and (\ref{eq:tilted_path_difference}) are fully consistent. First, one proves that $|{\bf q}| = 2 k \sin(\theta/2)$ and the angle between vectors $\bf q$ and $\delta {\bf R}$ equals $\alpha - \theta/2$. Then the scalar product is then given by
$$
{\bf q} \cdot \delta{\bf R} = |{\bf q}| |\delta {\bf R}| \cos\left(\alpha - \frac{\theta}{2}\right) = 2 k d \sin\left(\frac{\theta}{2}\right) \cos\left(\alpha - \frac{\theta}{2}\right)
$$
Performing a series of transformations
\begin{eqnarray*}
&& 2 \sin\left( \theta/2 \right) \cos\left(\alpha - \theta/2 \right)
\\
&=& 2 \sin\left(\theta/2 \right) \left[
\cos\alpha \cos\left( \theta/2 \right) + \sin\alpha \sin\left( \theta/2 \right)
\right]
\\
&=& \sin\theta \cos\alpha + 2 \sin^2\left( \theta/2 \right) \sin\alpha
\\
&=& \sin\theta \cos\alpha + (1 - \cos\theta)\sin\alpha
\\
&=& \sin(\theta-\alpha) + \sin\alpha = f(\theta,\alpha) \ ,
\end{eqnarray*}
we show the sought equivalence.

\begin{figure}[h]
\center
\includegraphics[width = 0.2\textwidth]{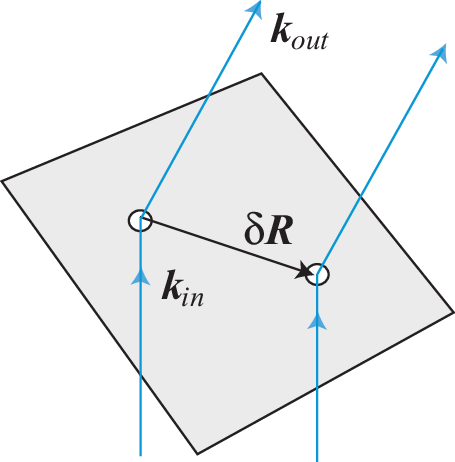}
\caption{Formula (\ref{eq:phase_in_vector_notation}) works in a general 3D setting}
 \label{fig:two_slits_3D}
\end{figure}

The advantage of formula (\ref{eq:phase_in_vector_notation}) comes from the fact that its derivation never assumes that vectors ${\bf k}_{in}$, ${\bf k}_{out}$, $\delta {\bf R}$ lie in the same plane. In fact, they can form an arbitrary configuration in 3D space. This means, for example, that expression (\ref{eq:phase_in_vector_notation}) can be readily used to describe diffraction on 2D gratings (Fig.~\ref{fig:two_slits_3D}), where finding $\Delta l$ via a purely geometrical approach would consitute a formidable task. Furthermore, (\ref{eq:phase_in_vector_notation}) can be as well used to describe the X-ray diffraction in 3D crystals.

\subsection{Reciprocal lattice}

\begin{figure}[b]
\center
\includegraphics[width = 0.4\textwidth]{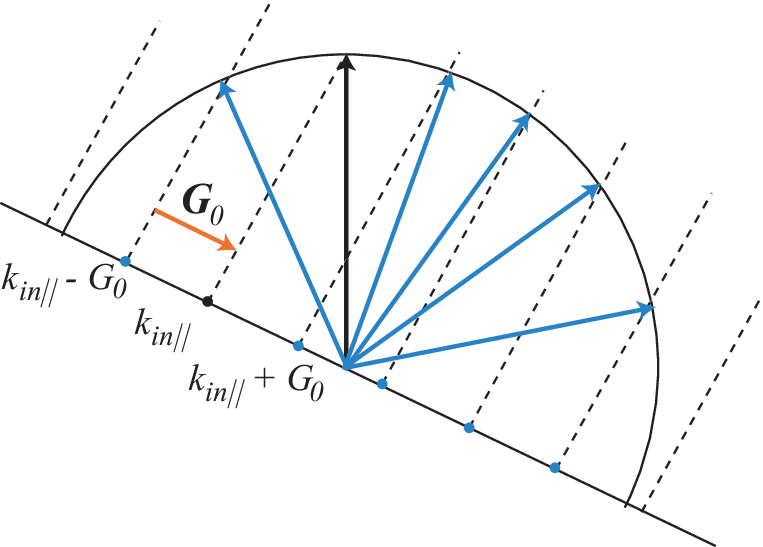}
\caption{Diffraction on a tilted grating from the reciprocal lattice perspective. Black vertical arrow is the incoming wave vector ${\bf k}_{in}$. The fan of blue arrows are the ${\bf k}_{out}$ vectors of the diffraction maxima. Points on the line of the grating are the reciprocal lattice. All elements of the figure belong to the space of wave vectors, as opposed to the real space.}
 \label{fig:reciprocal_lattice_1D}
\end{figure}

Let's now go back to the issue of main peak angles in the diffraction on a tilted grating. Using formula (\ref{eq:phase_in_vector_notation}) we can re-write condition (\ref{eq:main_maxima_locations}) as
$$ 
{\bf q} \cdot {\bf d} = 2\pi m \ ,
$$ 
where vector $\bf d$ is directed along the grating and has a length equal to the grating's period $d$. Denote $q_{||}$ to be the component of $\bf q$ along the grating. Then ${\bf q} \cdot {\bf d} = q_{||} d = (k_{out ||} - k_{in ||}) d$, where $k_{in,out ||}$ are the parallel components of ${\bf k}_{in,out}$. Peak condition can be now written as
\begin{equation}\label{eq:projection_condition}
k_{out ||} = k_{in ||} + \frac{2\pi m}{d}.
\end{equation}
In this form the condition states that at a diffraction peak the parallel component of the outgoing wave vector differs from the parallel component of the incoming wave vector by one of the numbers $G_m = G_0 m = 2\pi m/d$. Fig.~\ref{fig:reciprocal_lattice_1D} shows how that condition, together with the requirement $|{\bf k}_{in}| = |{\bf k}_{out}| = k$, determines all possible wave vectors ${\bf k}_{out}$ corresponding to the peaks. In this figure the blue points on the line of grating are positioned at equal distances $G_0$. They are obtained by shifting the black point that represents $k_{in ||}$ along the line of the grating with vectors ${\bf G}_0 m$, where
\begin{equation}\label{eq:reciprocal_lattice_period}
{\bf G}_0 = \frac{2\pi{\bf d}}{d^2} \ .
\end{equation}
The system of blue points is the {\em reciprocal lattice} associated with the grating. Dashed lines start at the reciprocal lattice points and are drawn perpenidicular to the line of the grating. Vectors ${\bf k}_{out}$ end at the points of intersection of dashed lines and a circle of radius $k$. When $m$ becomes too large, positive or negative, the dashed lines do not inetersect the circlue any more. This way only a finite number of peaks is formed.

The angles of the diffraction maxima $\theta_m$ can be read from Fig.~\ref{fig:reciprocal_lattice_1D}. In terms of the actual calculations, doing this is equivalent to inverting Eq.~(\ref{eq:main_maxima_locations}). In contrast, the reciprocal lattice approach provides an intuitive picture. For example, it allows one to predict qualitatively how the diffraction pattern changes when the tilt angle is varied. Imagine that one increases the tilt angle in Fig.~\ref{fig:reciprocal_lattice_1D}. The black projection point of ${\bf k}_{in}$ will shift up and left along the line of grating. The whole reciprocal lattice will follow it. This will lead to the disappearance of maxima on the left of ${\bf k}_{in}$ and appearance of new maxima on the right.

\subsection{A preview of diffraction on 2D and 3D periodic structures}
The 1D reciprocal lattice discussed here can serve as a stepping stone helping to understand the more complicated reciprocal lattices associated with 2D and 3D periodic structures.

2D diffraction gratings are commonly used in the classroom demonstration experiments (see, e.g., Refs.~\onlinecite{mak:PT1994:2D}, \onlinecite{logiurato:PT2020:2D}, and Sec. 10.2.8 of Ref.~\onlinecite{hecht_book}). In them, the pattern of apertures has two vectors of periodicity ${\bf d}_1$ and ${\bf d}_2$, both lying in the plane of the grating. Those vectors may be orthogonal to each other, as in a square or rectangular gratings, or have other angles between them, as in a triangular or  honeycomb gratings. The notion of reciprocal lattice has a straightforward generalization for two dimensions. Without going into the details, we list the similarities between the 1D and 2D cases. Two-dimensional reciprocal lattice is a 2D array of points lying in the grating plane. It has two vectors of periodicity ${\bf G}_{1}$ and ${\bf G}_{2}$, and condition (\ref{eq:projection_condition}) generalizes to
$$
{\bf k}_{out||} = {\bf k}_{in||} + m_1 {\bf G}_1 + m_2 {\bf G}_2 \ ,
$$
where ${\bf k}_{||}$ are the projections of wave vectors $\bf k$ on the plane of the grating. Wave vectors for diffraction peaks are found via a straightforward generalization of  Fig.~\ref{fig:reciprocal_lattice_1D}. Instead of a half circle over the line, there is a half-sphere over the plane. Dashed lines, starting at each reciprocal lattice point, extend perpendicularly to the plane, until they cross the sphere. Wave vectors ${\bf k}_{out}$ corresponding to the diffraction peaks start at the origin and end in those crossing points. One important difference between the 1D and 2D gratings is that the relation between vectors ${\bf G}_{1,2}$ and ${\bf d}_{1,2}$ is more complicated \cite{kittel-book}. We will not discuss it here.

3D periodic diffraction gratings are most commonly realized by crystals, where atoms are naturally arranged periodically in space. When crystals are placed into a beam of X-rays,
each atom 
serves as a secondary source. Those sources are arranged into a 3D periodic lattice with periods $({\bf d}_1, {\bf d}_2, {\bf d}_3)$.

The physics of scattering in 3D has one important difference from the diffraction on 1D and 2D gratings: in order to excite a given secondary source, the light has to travel to it through the crystal itself. Note that in the gratings considered before there were no obstacles between the incoming beam and the secondary sources. Moreover, waves emitted by a secondary sources were propagating freely afterwards, while in a crystal they have to a travel through the sample before reaching the free space.

Fortunately, these complications are sometimes not important. The mathematics of crystal diffraction remains si\-mi\-lar to that of 1D and 2D gratings in the limit of weak scattering, that is when light almost completely passes through the crystal, and only its tiny fraction of it is diffracted. While it is difficult to achieve such situation for the visible light, X-rays do operate in the regime of weak scattering. The same can be true for the microwaves send through the artificial periodic structures
\cite{allen:AJP1955:microwaves, murray:AJP1974:microwaves, amato:AJP2009:microwaves}. In such cases diffraction maxima (also called ``reflexes'') are associated with wave vectors obeying the 3D generalization of condition (\ref{eq:projection_condition})
$$
{\bf k}_{out} = {\bf k}_{in} + m_1 {\bf G}_1 + m_2 {\bf G}_2 + m_3 {\bf G}_3
$$
(no need to perform projections). The generalization of Fig.~\ref{fig:reciprocal_lattice_1D} to the case of 3D crystal diffraction is called the ``Ewald's sphere construction'' \cite{kittel-book}.  Just as in the 2D case, we leave out the discussion of the procedure determining the reciprocal lattice periodicity vectors $({\bf G}_1, {\bf G}_2, {\bf G}_3)$ but refer the reader to Ref.~\onlinecite{kittel-book}.

\section{Disordered diffraction gratings}\label{sec:disordered_gratings}
Standard kits for diffraction demonstration experiments include a slide with randomly positioned circular apertures. The number of those apertures is the same as on another slide where they are arranged periodically, nevertheless the diffraction patterns obtained from the two slides are drastically different. Slide with periodically arranged apertures produces bright points from major diffraction peaks, while the aperiodic slide creates a picture generally resembling the diffraction pattern from a single aperture. In this section we will numerically model that demonstration experiment and show how the Huygens-Fresnel approach is able to reproduce not just the rough picture of diffraction but also some non-obvious details of the pattern.

\subsection{The issue with averaging over random phases}
Diffraction pattern produced by randomly positioned apertures is usually explained in the following way. As we know from Sec.~\ref{sec:form-factor}, the total intensity is a product of the grating factor and form factor (note that periodicity was not essential for deriving that result: it remains valid as long as all apertures are identical). The grating factor is responsible for the narrow intensity peaks observed for periodic arrangement of the apertures. Let's see how it behaves when the aperture positions are random. We need to compute
$$
I_g(\theta) = \left|
\sum_{p = 1}^{N} e^{- i {\bf q} \cdot \delta {\bf R}_p }
\right|^2
$$
with random vectors $\delta {\bf R}_p$. (In the case of periodic grating we had $I_g = I_N/I_0$).  Re-write this as
\begin{eqnarray*}
&& \left(
\sum_{p = 1}^{N} e^{- i {\bf q} \cdot \delta {\bf R}_p }
\right) \times
\left(
\sum_{m = 1}^{N} e^{i {\bf q} \cdot \delta {\bf R}_m }
\right) =
\sum_{p, m} e^{- i {\bf q} \cdot (\delta {\bf R}_p - \delta {\bf R}_m) }
\\
&&
\qquad = N + \sum_{p \neq m} e^{- i {\bf q} \cdot (\delta {\bf R}_p - \delta {\bf R}_m) }
\end{eqnarray*}

\begin{figure}[t]
\center
\includegraphics[width = 0.25\textwidth]{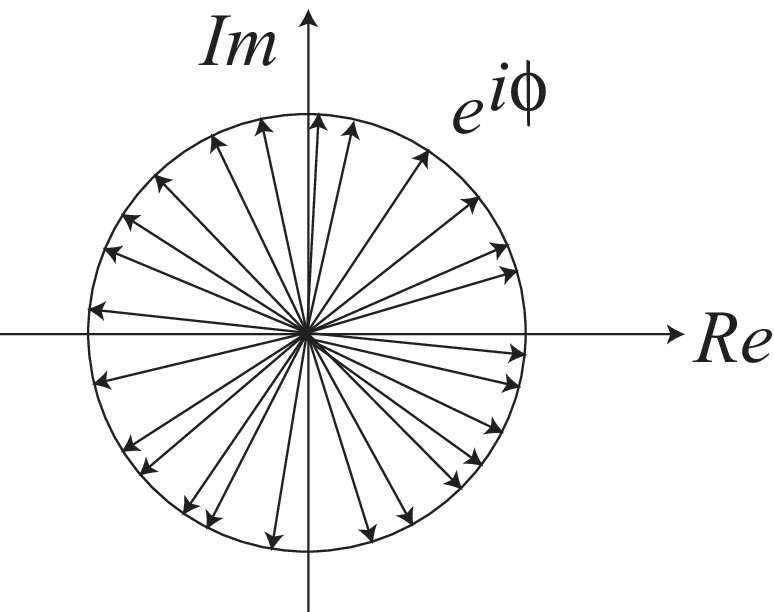}
\caption{Random phase factors $e^{i\phi}$ as vectors on the complex plane}
 \label{fig:random_phase_summation}
\end{figure}

The argument then goes to say that the second term is a sum of many complex numbers of the type $e^{i\phi}$ with random phases $\phi$. Figure~\ref{fig:random_phase_summation} suggests that on average such sum must be equal zero. Indeed finding this sum is equivalent to adding vectors shown in the figure. The length of each vector is unity, while the direction is random. Thus there are lots of almost antiparallel vectors that should almost cancel each other. In the limit of $N \to \infty$ this cancellation should become perfect, giving
\begin{equation}\label{eq:Ig_random_grating_naive}
I_g \to N \ .
\end{equation}

\begin{figure}[t]
\center
\includegraphics[width = 0.4\textwidth]{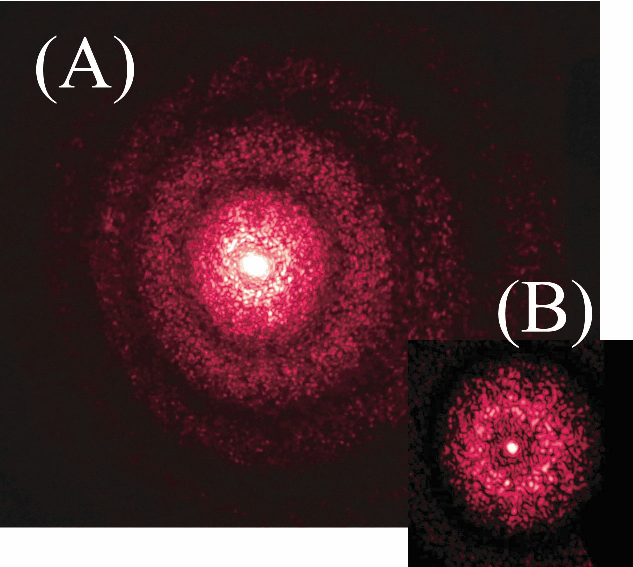}
\caption{Photograph of an experimental diffraction pattern on randomly positioned apertures. (A) larger pattern of rings corresponding to the form factor. Central part is overexposed and does not show the details; (B) Detailed view of the central part: bright spot at zero diffraction agle, the intensity depression around it, and the speckles }
 \label{fig:random_apertures_exp}
\end{figure}

The conclusion is that a random grating factor is simply a constant equal to $N$, independently of the wave vector $\bf q$. The total diffraction intensity is therefore the form factor multiplied by the number of apertures. This result is summarized by saying that for randomly positioned apertures the intensities are summed, while the interference terms average to zero.

An actual photograph of the random grating diffraction pattern is shown in Fig~\ref{fig:random_apertures_exp}. One can see that it does not exactly follow the predictions discussed above. Instead of the smooth intensity variation expected from a single circular aperture (Ref.~\onlinecite{hecht_book}, Sec. 10.2.5), there is a sharp bright spot in the center, surrounded by a darker ring. Another feature of the image are the irregular variation of intensity, known as {\em speckles}, across the whole pattern.

Following our general approach, we will use numeric modeling to check whether those features can be reproduced by the Huygens-Fresnel theory.

\subsection{1D random grating}
Let us start from a conceptually simpler case of a 1D diffraction grating. Actual exerimental studies of that case were performed in Ref.~\onlinecite{licinio:PT1999:random_gratings}.

The first thing to realize is that a ``grating with random disorder'' is a term requiring further clarification. One needs to precisely specify the type of disorder to be realized. Some features of diffraction on random gratings may be insensitive to the type of disorder but others will vary from one disorder to another. For a grating with $N$ identical, finite-size slits the description of disorder is given in the form of the probability $P(x_1, x_2, \ldots x_N)$ for the slits to be located at positions $(x_1, x_2, \ldots x_N)$. An actual physical grating has the slits' positions fixed once and for all. It is a {\em realization} of disorder.

We will employ the following simple model of disorder. Start with a perfectly periodic grating of slits repeating with a period $d$. After that, displace every slit by a random distance $u$ that is uniformly distributed in an interval $(-u_0,u_0)$. Position of the slit number $p$ is given by $x_p = p d + u_p$.  A realization of disorder is given by a set of displacements $(u_1, u_2, \ldots , u_N)$ which are assumed to be independent of each other, meaning
\begin{equation}\label{eq:probability_of_randomness_ensemble}
P(u_1, u_2, \ldots u_N) = p(u_1) p(u_2) \ldots p(u_N)
\end{equation}
with
\begin{equation}\label{eq:probability_of_randomness_one_site}
p(u) = \frac{ \theta(u + u_0) - \theta(u - u_0) }{2u_0}.
\end{equation}
The disorder is weak for $u_0 \ll d$.

Assuming that the incoming light beam falls normally (no tilting), the grating factor equals
\begin{equation}
I_g(\theta) = \left|
  \sum_{p = 1}^{N} e^{i k (p d + u_p)  \sin\theta}
\right|^2 ,
\end{equation}
and the full intensity is then given by
$$
I(\theta) = I_g(\theta) J(\theta) \ .
$$

\begin{figure}[t]
\center
\includegraphics[width = 0.39\textwidth]{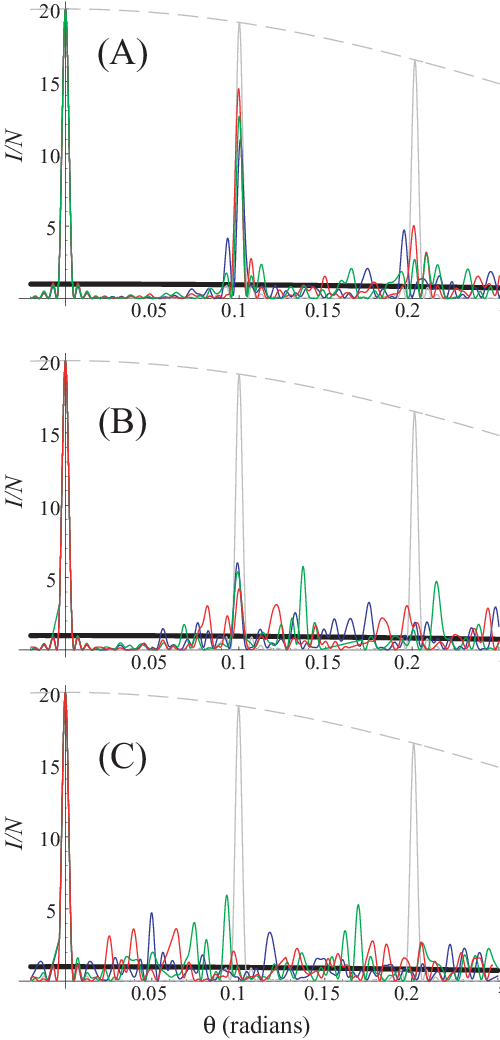}
\caption{Diffraction patterns of 1D gratings with disorder (normalized intensity $I/N$). Parameter $u_0/d$ takes values (A) $0.2$, (B) $0.3$, and (C) $0.95$. Other parameters are $N = 20$, $w = 1.2 \lambda$, $d = 10 \lambda$. Tall gray peaks are the normalized intensity pattern $I_N/N$ of an exactly periodic grating. Thick black line is the form factor of a single slit $J(\theta)$. Three colored lines correspond to three realizations of disorder.}
 \label{fig:random_grating}
\end{figure}

Figure~\ref{fig:random_grating} shows the results of simulating $I(\theta)$ with a progressively increasing disorder. One sees that the increase of $u_0$ produces three outcomes.

First, all major peaks, except for the central one, gradually decrease in height and assume random magnitudes. The central peak is unaltered by the disorder.

Second, noisy intensity fluctuations develop between the major peaks. Fluctuations' heights are much larger than what was observed for a perfectly periodic grating. For some angles intensity goes up much higher than the line $J(\theta)$ predicted by the naive averaging, while for others it drops down to zero.

Third, for relatively small disorder the fluctuations do not develop in the vicinity of the central peak. The size of the fluctuation-free interval is decreasing with increasing $u_0$, until for $u_0/d \to 1$ all space around the central peak is filled with fluctuations and looks no different from the other angular intervals.

The three features discussed above correlate well with the experimentally observed pattern shown in Fig.~\ref{fig:random_apertures_exp}. It has a bright central peak, a depression of intensity around it, and the speckles seen in the photograph are nothing else but the manifestation of the fluctuations of intensity predicted by the Huygens-Fresnel simulation.

The first two features can be quite easily justified analytically. The robustness of the central peak comes from the fact that at $\theta = 0$ all phase factor are equal to unity, independent of the disorder realization. Thus the height of the central peak remains equal to $I_g(0) = N^2$. For a nonzero but small angle, one can approximate
$$
 e^{i k (p d + u_p)\sin\theta} =  e^{i k p d\sin\theta} (1 + k u_p \sin\theta + \ldots )
 \approx e^{i k p d\sin\theta}
$$
as long as the condition $k u_0 \sin\theta \ll 1$ holds. That explains why the diffraction intensity at the central peak, and in an angular interval $|\theta| \lesssim 1/(k u_0)$ around it, remains close to that of a perfectly periodic grating.

The third feature, namely the existence of the large intensity fluctuations producing the speckles, will require a more serious mathematical approach. Nevertheless, numeric simulations convince us that it is a robust feature of diffraction on random gratings.

A side note about the intensity fluctuations should be made here. It is natural to ask whether averaging them over a finite angle window, for example as
\begin{equation}\label{eq:angle_averaging}
\langle I(\theta) \rangle = \frac{1}{2\Delta\theta}
\int_{\theta - \Delta\theta}^{\theta+\Delta\theta} I(\theta') d\theta' \ ,
\end{equation}
would result in recovering the smooth single-aperture intensity (\ref{eq:Fraunhofer_intensity_finite-width_analytic}). This is a perfectly legitimate question but we want to underscore that it does not reduce to a straightforward averaging over the probability distributions (\ref{eq:probability_of_randomness_ensemble}, \ref{eq:probability_of_randomness_one_site}). In expression (\ref{eq:angle_averaging}) the integrand is a random variable taken for a given realization of disorder. So the right hand side $\langle I(\theta) \rangle$ is also a random variable, as opposed to a number. The meaningful question to ask is about the probability distribution of $\langle I(\theta) \rangle$ given the underlying probability distribution of disorder. Discussion of that matter is beyond the scope of our paper.

Overall, one sees that while the naive averaging (\ref{eq:Ig_random_grating_naive}) does not tell the whole story, a consistent application of the Huygens-Fresnel theory explains experimental findings quite well.

\subsection{2D random grating}
We now perform a 2D simulation to make sure that it produces the same features as found from the 1D simulation, and that it corresponds to the 2D experimental results shown in Fig.~\ref{fig:random_apertures_exp}.

We compute the grating factor for a collection of randomly positioned apertures on a 2D screen. The model of randomness in 2D is a straightforward generalization of the one used in the 1D case. Positions of the apertures are given by vectors
$$
r_{pq} = (p d + u^x_{pq}, q d + u^y_{pq})
$$
that are obtained from a square lattice with period $d$, by shifting each aperture by a random vector ${\bf u}_{ij}$. Components of ${\bf u}_{ij}$ are independent random variables with probability distribution (\ref{eq:probability_of_randomness_ensemble}, \ref{eq:probability_of_randomness_one_site}).

\begin{figure}[t]
\center
\includegraphics[width = 0.4\textwidth]{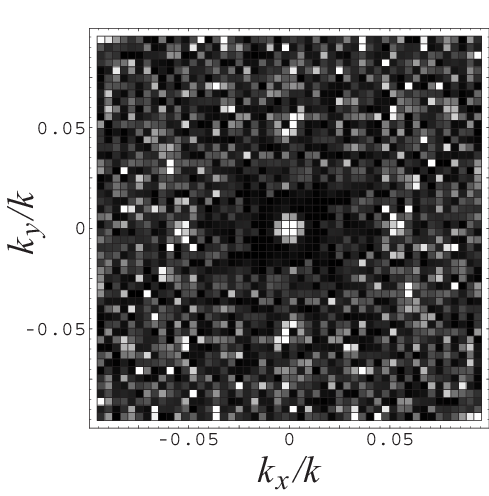}
\caption{Density plot for the 2D grating factor vs. the components of the outgoing wave vector ${\bf k}_{out}$ with $N = 144$ apertures, $d = 18 \lambda$, and $u_0 = 7.2 \lambda$. Incoming beam is perpendicular to the grating.}
 \label{fig:2D_random}
\end{figure}

Results of the simulations are shown in Fig.~\ref{fig:2D_random}. The central peak, the intensity depression around it, and the pattern of speckles are all successfully reproduced.

\section{Discussion}
Numeric simulations are most useful in a teaching context when they are discussed together with the demonstration experiments. Their role is to provide a framework that allows students to understand the observed patterns without performing careful and time-consuming measurements.

Simulations for gratings with $N$ slits discussed in Sec.~\ref{sec:H-F} should be compared to diffraction patterns obtained from sequences of slides starting with a two-slit slide and going up to 5--6 slits. Such sequences are included in most demonstration kits.

Slides with hundreds of slits should be discussed based on the material of Sec.~\ref{sec:reciprocal_lattice}. The patterns produced by them on the screen are essentially the reciprocal lattices and students can observe how their periods obey the relation (\ref{eq:reciprocal_lattice_period}).

Finally, slides with random apertures together with material from Sec.~\ref{sec:disordered_gratings} give an engaging introduction into the peculiar behavior of coherent light beyond the domain of periodic structures.

\acknowledgments
The author wishes to thank TeachSpin, Inc. and the Jonathan F. Reichert Foundation for supporting his participation in the Conference on the Integration of Computation, Experiment, and Physical Theory, where this work was started. Many thanks go to Mr. James Clawson for help with the experimental setup.


\end{document}